\documentclass{lhep}       

\journal{LHEP}%Letters in High Energy Physics}
\jyear{2023}
\pages{NuDM-2022 } 

\received{15 November 2022}
\published{xx March 2018}

\def\be{\begin{equation}}
\def\ee{\end{equation}}
\def\bea{\begin{eqnarray}}
\def\eea{\end{eqnarray}}

\usepackage[T1]{fontenc}
\usepackage[english]{babel}
\usepackage{lmodern}
\usepackage[colorlinks = true]{hyperref}
\usepackage{amsfonts}
\usepackage{amssymb}
\usepackage{bbm}
\usepackage{bm}
\usepackage{color}
\usepackage{slashed}
\usepackage{graphicx}
\usepackage{dsfont}
\usepackage{bbold}
\usepackage{braket}

\begin{document}

\title{Effective Field Theories for Dark Matter Pairs in the Early Universe}

\author{S.~Biondini,\auno{1} N.~Brambilla,\auno{2,3} G.~Qerimi,\auno{2} and A. Vairo\auno{2}}
\address{$^1$Department of Physics, University of Basel, Klingelbergstr. 82, CH-4056 Basel, Switzerland}
\address{$^2$Physik-Department, Technical University Munich, James-Franck-Str.  1, 85748 Garching, Germany}
\address{$^3$Institute for Advanced Study, Technical University Munich, Lichtenbergstrasse 2 a, 85748 Garching, Germany}

\begin{abstract}
In this conference paper, we consider effective field theories of non-relativistic dark matter particles interacting with a light force mediator in the early expanding universe. We present a general framework, where to account in a systematic way for the relevant processes that may affect the dynamics during thermal freeze-out. In the temperature regime where near-threshold effects, most notably the formation of bound states and Sommerfeld enhancement, have a large impact on the dark matter relic density, we scrutinize possible contributions from higher excited states and radiative corrections in the annihilations and decays of dark-matter pairs.
\end{abstract}

\maketitle


\section{Introduction}
Complementary astrophysical observations strongly support the evidence that 80\% of the present matter consists of dark matter (DM), and anisotropy measurements in the CMB determine precisely its relic density to be $\Omega_{\hbox{\tiny DM}} h^2 = 0.1200 \pm 0.0012$.~\cite{Akrami:2018vks} Despite its nature being elusive, an extensive work has been put forward and a variety of models have been constrained to reproduce this density~\cite{Bertone:2004pz,Feng:2010gw}. A prominent class of models involve heavy thermal DM particles, often referred to as Weakly Interacting Massive Particles (WIMPs), appearing in a typical thermal freeze-out scenario. In this proceeding, we focus on fermionic DM that experiences self-interactions through a long-range mediator within the dark sector. In particular, we consider a QED-like model and study the interactions within a thermal bath of dark photons. The proceeding is based on ref.~\cite{B_and_NAG_1}. In section~\ref{sec:pNREFT} we show how to construct the DM non-relativistic effective field theories (EFTs). In section~\ref{sec:processes}, we compute, in the EFTs, near threshold observables. We give the DM abundance by solving the rate equations in section~\ref{sec:rateequations} upon including the relevant processes. Finally, conclusions and outlook are in section~\ref{sec:concl}.

\section{NREFT$_{\textrm{DM}}$}%{{ WHAT IS \LaTeX\,?}}
\label{sec:pNREFT}
The Lagrangian density of a dark Dirac fermion $X$ charged under an abelian gauge group reads
\begin{equation}
\mathcal{L}=\bar{X} (i \slashed {D} -M) X -\frac{1}{4} F_{\mu \nu} F^{\mu \nu} + \mathcal{L}_{\textrm{portal}} \, ,
\label{lag_mod_0}
\end{equation}
where the covariant derivative is $D_\mu=\partial_\mu + i g A_\mu$, with $A_\mu$ the dark photon field and $F_{\mu \nu} = \partial_\mu A_\nu - \partial_\nu A_\mu$. The dark fine structure constant is $\alpha \equiv g^2/(4 \pi)$.
Additional interactions coupling the dark photon with the SM degrees of freedom (d.o.f.), such as kinetic mixing~\cite{Holdom:1985ag,Foot:1991kb}, are comprised in $\mathcal{L}_{\textrm{portal}}$, which are beyond the scope of this work and thus are omitted.

We are interested in processes close to threshold, i.e., processes involving pairs of non-relativistic dark fermions with relative velocities $v_{\text{rel}} \sim \alpha \ll 1$. The dark photons form a thermal bath of temperature $T$ that is weakly coupled to the DM. If the latter is thermalized then DM momenta scale like $p \sim \sqrt{MT}$. The scales are assumed to be hierarchically ordered as
\begin{equation}
M \gg M\alpha \gg M\alpha^2 \gtrsim T \,.
\label{scale_arrang}
\end{equation}
In a typical freeze-out scenario the decoupling from chemical equilibrium happens around $M/T \approx 25$. The clear separation of different scales in \eqref{scale_arrang} allows to build a tower of EFTs starting from \eqref{lag_mod_0}, and extract the relevant interactions and corresponding observables around the decoupling time. Near threshold effects comprise the annihilation of DM pairs as well as electric transitions within the pairs. Higher multipole transitions will be suppressed at later times, i.e., smaller $T$. These processes play an important role for a quantitative treatment of the dynamics of the relevant d.o.f. in the early universe, and the corresponding observables appear in the evolution equations.

Integrating out hard modes leads to a non-relativistic EFT, here dubbed NRQED$_{\textrm{DM}}$ ~\cite{Caswell:1985ui}. Hard processes such as annihilations, happening at a scale $\sim M$, are encoded at leading order in the non-relativistic expansion in the matching coefficients of dimension-6 four-fermion operators that overlap only with S-waves.\footnote{P-wave annihilations start contributing from dimension-8 four-fermion operators, and therefore are suppressed at low energies.} At order $\alpha^3$ their imaginary parts read~\cite{Barbieri:1979be,Hagiwara:1980nv}
\begin{eqnarray}
  % &&  {\rm{Im}} (d_s) = \pi \alpha(\mu_{\hbox{\tiny M}})^2 \left[ 1-\frac{\alpha(\mu_{\hbox{\tiny M}})}{\pi} \left( 5-\frac{\pi^2}{4} \right)\right] \, ,
           &&  {\rm{Im}} (d_s) = \pi \alpha^2 \left[ 1+\frac{\alpha}{\pi} \left(\frac{\pi^2}{4} -5\right)\right] \, ,
    \label{Im_ds_NLO}
      \\
      % &&  {\rm{Im}} (d_v) = \frac{4}{9} (\pi^2 -9) \alpha(\mu_{\hbox{\tiny M}})^3 \, ,
               &&  {\rm{Im}} (d_v) = \frac{4}{9} (\pi^2 -9) \alpha^3 \, .
    \label{Im_dv_NLO}
   \end{eqnarray}
They originate from S-wave spin-singlet ($X \bar{X} \to \gamma \gamma$) and spin-triplet ($X \bar{X} \to \gamma \gamma \gamma)$ annihilations, respectively.

Next we integrate out modes associated to the soft scale $M\alpha$. In order to enforce that the photon fields do not depend on the soft scale anymore, they are multipole expanded in the relative coordinate $\bm{r} \equiv \bm{x}_1-\bm{x}_2$ of the pair, i.e., the distance between a fermion located at $\bm{x}_1$ and an antifermion located at $\bm{x}_2$. The effective field theory has the form of potential NRQED (pNRQED)~\cite{Pineda_1998,Pineda:1997ie} and we denote it as pNRQED$_\textrm{DM}$. Its Lagrangian is given by
\begin{equation}
\begin{aligned}
  \mathcal{L}_{\textrm{pNRQED}_{\textrm{DM}}}&=   \int d^3 \bm{r} \phi^\dagger(t,\bm{r},\bm{R})
              [ i \partial_0 -H(\bm{r},\bm{p},\bm{P},\bm{S}_1,\bm{S}_2) \\
&~~~~+ g \, \bm{r} \cdot \bm{E}(t,\bm{R})] \phi (t,\bm{r},\bm{R}) -\frac{1}{4} F_{\mu \nu} F^{\mu \nu}
 \, ,
\label{pNREFT_1}
\end{aligned}
\end{equation}
where 
\begin{equation}
\begin{aligned}
 H(\bm{r},\bm{p},\bm{P},\bm{S}_1,\bm{S}_2) &= 2M + \frac{\bm{p}^2}{M}+\frac{\bm{P}^2}{4M} - \frac{\bm{p}^4}{4M^3} \\
 &~~~+  V (\bm{r},\bm{p},\bm{P},\bm{S}_1,\bm{S}_2) + \ldots \, ,
 \label{ham_pNRQED}
\end{aligned}
\end{equation}
\begin{equation}
\begin{aligned}
  &V (\bm{r},\bm{p},\bm{P},\bm{S}_1,\bm{S}_2)= V^{(0)} + \frac{V^{(1)}}{M} + \frac{V^{(2)}}{M^2} + \ldots \, ,
 \label{pot_pNRQED}    
\end{aligned}
\end{equation}
and $\bm{S}_1=\bm{\sigma}_1/2$ and $\bm{S}_2=\bm{\sigma}_2/2$ are the spin operators acting on the fermion and antifermion, respectively.
At leading order the static potential is the Coulomb potential $V^{(0)}=-\alpha/r$. While the photon field depends only on the center-of-mass (c.o.m.) position $\bm{R} \equiv (\bm{x}_1+\bm{x}_2)/2$, the bilocal field of the dark pair depends on both $\bm{r},\bm{R}$ and can be decomposed into a scattering and bound state part~\cite{Yao:2018nmy}
\begin{equation}
\begin{aligned}
  \phi_{ij}(t,\bm{r},\bm{R}) &= \int \frac{d^3\bm{P}}{(2\pi)^3} \Bigg[ \sum_{n,\textrm{spin}} e^{-iE_nt+i\bm{P}\cdot\bm{R}}\,\Psi_n(\bm{r})\,S_{ij}\,\phi_{n}(\bm{P})
                           \\
  &\hspace{0.3cm}  + \sum_{\textrm{spin}}\int \frac{d^3\bm{p}}{(2\pi)^3} e^{-iE_pt+i\bm{P}\cdot\bm{R}}\,\Psi_{\bm{p}}(\bm{r})\,S_{ij}\,\phi_{p}(\bm{P})\Bigg] \,.
\end{aligned}
\end{equation}
DM annihilations, inherited from the imaginary parts of the NRQED$_\textrm{DM}$ matching coefficients, induce the following local potential
\begin{equation}
\begin{aligned}
  \delta V^{\textrm{ann}}(\bm{r})
  &= - \frac{i \delta^3(\bm{r})}{M^2}  \left[ {2\rm{Im}}(d_s) - \bm{S}^2 \left( {\rm{Im}}(d_s)- {\rm{Im}}(d_v) \right) \right] \\
  &~~~~+ \dots\;,
\label{pNREFT_2}
\end{aligned}
\end{equation}
where $\bm{S}=\bm{S}_1+\bm{S}_2$ is the total spin of the pair and the dots comprise annihilations with non-vanishing orbital angular momentum.
The case of pNRQED at finite temperature has been studied in refs.~\cite{Escobedo:2008sy,Escobedo:2010tu}, whereas an application to DM models with scalar mediators can be found in refs.~\cite{Biondini:2021ccr,Biondini:2021ycj}. The matching is done in the weakly coupled regime order by order in $\alpha$, although the EFT is suited to accommodate a non-perturbative framework as well~\cite{Brambilla:1999xf,Brambilla:2004jw}. The dynamics at the soft scale is encoded in the matching coefficients of pNRQED$_{\textrm{DM}}$ which are the potentials. The equations of motion are of the Schrödinger type, where the potentials distort the free wavefunctions into bound-state wavefunctions $\Psi_n(\bm{r})\,S_{ij}$ with discrete energies $E_n$, or into scattering wavefunctions  $\Psi_{\bm{p}}(\bm{r})\,S_{ij}$ with positive energies $E_p$.\footnote{The spin wavefunction $S_{ij}$ accounts for the pairs being either in a spin-singlet or spin-triplet configuration, $E_n=2M-M\alpha^2/(4n^2) + \dots$ and $E_p=2M+p^2/M + \dots ~= 2M+Mv_{\textrm{rel}}^2/4 + \dots$ .}

By exploiting EFT techniques to separate the various scales being initially intertwined in \eqref{lag_mod_0}, we arrive at a thermal field theory that describes dark fermion-antifermion pairs and dark photons of energy of the order of or below the Coulomb binding energy.

\section{Near-threshold processes} 
\label{sec:processes}
Though pair-annihilation is a process appearing at the hard momentum scale, it encompasses near-threshold effects induced by repeated soft dark photon exchanges. Resumming such multiple rescatterings for dark fermion-antifermion pairs above threshold, i.e. when they form a scattering state, results in a Sommerfeld enhanced spin-averaged S-wave annihilation cross section
\begin{equation}
\begin{aligned}
(\sigma_{\hbox{\scriptsize ann}} v_{\hbox{\scriptsize rel}})(\bm{p}) 
&= \frac{{\rm{Im}}(d_s)+3{\rm{Im}}(d_v)}{M^2} \left|\Psi_{\bm{p}}(\bm{0})\right|^2 \\
    &=  (\sigma^{\hbox{\tiny NR}}_{\hbox{\scriptsize ann}} v_{\hbox{\scriptsize rel}})  \, S_{\hbox{\scriptsize ann}}(\zeta)\,,
\label{ann_fact_scat}
\end{aligned}
\end{equation}
in the c.o.m. frame. The velocity-independent contribution from the hard scale is separated from the soft-scale dependent Sommerfeld factor (for S-waves)~\cite{Sommerfeld,Cassel:2009wt}
\begin{equation}
S_{\hbox{\scriptsize ann}}(\zeta)=\left|\Psi_{\bm{p}}(\bm{0})\right|^2=\frac{2 \pi \zeta}{1-e^{-2 \pi \zeta}} \, ,
\label{Somme_0}
\end{equation}
with $\zeta \equiv \alpha/v_{\hbox{\scriptsize rel}}$ and $p = Mv_{\hbox{\scriptsize rel}}/2$. On the other hand, for pairs below threshold, the relevant observables for annihilation are the decay widths given by
\begin{equation}
\Gamma^{n, \hbox{\scriptsize para}}_{\hbox{\scriptsize ann}}= \frac{4 {\rm{Im}}(d_s)}{M^2} \left|\Psi_{n}(\bm{0})\right|^2 \,,
\label{ann_para}
\end{equation}
\begin{equation}
\Gamma^{n, \hbox{\scriptsize ortho}}_{\hbox{\scriptsize ann}}= \frac{4 {\rm{Im}}(d_v)}{M^2} \left|\Psi_{n}(\bm{0})\right|^2 \,,
\label{ann_ortho}
\end{equation}
for spin-singlet and spin-triplet S-wave bound states. We call them paradarkonium and orthodarkonium, respectively. We remark that the soft dark photon resummation effects are embedded already at the level of the Lagrangian~\eqref{pNREFT_1}, and that the annihilation rates follow directly from (-2) times the imaginary parts of the expectation values of~\eqref{pNREFT_2}.

\begin{figure}[t!]
    \centering
    \includegraphics[scale=0.33]{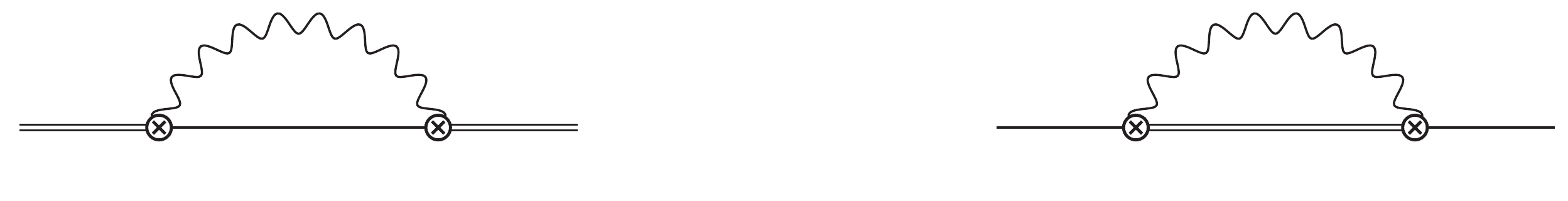}
    \caption{(Left) One-loop self-energy diagram of a scattering state (double solid line) in  pNRQED$_{\textrm{DM}}$; the imaginary part is related to the bound-state formation process $(X\bar{X})_p\to \gamma + (X\bar{X})_n$. (Right) One-loop self-energy diagram of a bound state (solid line); the imaginary part is responsible for the bound-state dissociation process $\gamma + (X\bar{X})_n\to (X\bar{X})_p$.
       The cross circled vertex denotes the electric dipole interaction in~\eqref{pNREFT_1}.}
    \label{fig:pnEFT_DM_self}
\end{figure}

Besides local interactions, the Lagrangian~\eqref{pNREFT_1} contains an electric dipole term that allows for the formation of a bound state through low-energy photon emission of a scattering state and viceversa, i.e., the dissociation of a bound state by absorption of a dark photon from the thermal bath. We abbreviate the processes by bsf and bsd, respectively. Their thermal rates may be computed, using the optical theorem, from the self-energy diagrams in pNRQED$_\textrm{DM}$, see fig.~\ref{fig:pnEFT_DM_self}.
Using the real-time formalism we obtain for the bsf cross section
\begin{equation}
\begin{aligned}
(\sigma_{\hbox{\scriptsize bsf}} \, v_{\hbox{\scriptsize rel}})(\bm{p})
  &= \frac{g^2}{3\pi}
  \sum_{n}
  \left[ 1 + n_{\text{B}}(\Delta E_{n}^{p}) \right]\left[ 1 + n_{\text{B}}(E_n) \right] \\
  &\hspace{2cm}\times |\langle \, n | \bm{r} | \bm{p} \,\rangle|^2
  (\Delta E_{n}^{p})^3 \, ,
\label{sigmabsf12}
\end{aligned}
\end{equation}
where $n_{\textrm{B}}$ is the Bose-Einstein distribution and $\Delta E_{n}^{p} = E_p-E_n = (Mv_{\textrm{rel}}^2/4)\left[1+\alpha^2/(n^2v_{\textrm{rel}}^2)\right] + \dots \;.$
The dots stand for higher order corrections in the energy spectrum. For the bsd width, we get the convolution integral
\begin{equation}
\begin{aligned}
     \Gamma^n_{\textrm{bsd}} =   2\int _{|\bm{k}|\geq |E_{n}|} \frac{d^{3}k}{(2\pi)^{3}}~n_{\textrm{B}}(|\bm{k}|)\left[ 1 + n_{\textrm{B}}(E_p) \right] \sigma^n_{\textrm{bsd}}(\bm{k}) \, ,
\label{gamma_diss_generic}
\end{aligned}
\end{equation}
where $2$ are the dark photon polarizations and $\sigma^n_{\textrm{bsd}}$ is the photo-dissociation cross section averaged over the photon polarizations, which reads
\begin{equation}
\begin{aligned}
     \sigma^n_{\textrm{bsd}}(\bm{k}) &= \frac{1}{2} \frac{g^{2}}{3\pi}\frac{M^{3/2}}{2}|\bm{k}|\sqrt{|\bm{k}|+E^b_{n}} \\
     &\hspace{2cm}\times \left| \left< n|\bm{r}|\bm{p}\right>\right|^{2}\bigg \vert_{|\bm{p}| = \sqrt{M(|\bm{k}|+E^b_{n})}} \, ;
     \label{diss_generic}
\end{aligned}
\end{equation}
$E_n^b$ is the binding energy: $E_n^b = E_n -2M$.

Further low-energy processes are the bound state-to-bound state de-excitation transitions, whose termal widths are
\begin{equation}
\begin{aligned}
  \Gamma^{n}_{\textrm{de-ex.}} &=
  \sum_{n'< n}\frac{g^2}{3 \pi}\left(\Delta E_{n'}^{n}\right)^3
  \left[1+n_{\text{B}}\left(\Delta E_{n'}^{n}\right)\right] \\
  &\hspace{2cm} \times \left[ 1 + n_{\text{B}}(E_{n'}) \right] \left| \langle \, n' \, |  \bm{r}  | \, n \, \rangle \right|^2 \, ,
    \label{gamma_deexcitation1}
\end{aligned}
\end{equation}
and similarly for excitations
    \begin{equation}
    \begin{aligned}
    \Gamma^{n}_{\textrm{ex.}} &= \sum_{n'> n}\frac{g^2}{3 \pi}\left|\Delta E_{n'}^{n}\right|^3  n_{\text{B}}\left(\left|\Delta E_{n'}^{n}\right|\right)\\
  &\hspace{2cm} \times \left[ 1 + n_{\text{B}}(E_{n'}) \right]\left| \langle \, n' \, |  \bm{r}  | \, n \, \rangle \right|^2 \, ,
    \label{gamma_excitation1}
\end{aligned}
\end{equation}
with $\Delta E_{n}^{n'} = E_{n'}-E_n = (M\alpha^2/4)\left(1/n^2-1/n'^2\right) + \dots \;.$
We observe that each of the rates in~\eqref{sigmabsf12},~\eqref{gamma_diss_generic},~\eqref{gamma_deexcitation1} and~\eqref{gamma_excitation1} factorizes into a thermal part and an in-vacuum part involving the electric dipole matrix element squared. The thermal component can be further simplified, since the distribution functions $n_{\text{B}}(E_{n}), n_{\text{B}}(E_{p})$ vanish exponentially for the temperature region set by~\eqref{scale_arrang}. In that limit our result for the bsf agrees with the outcomes in~\cite{Petraki:2015hla,Binder:2020efn} and for bsd with the ones in~\cite{Escobedo:2008sy} for the hydrogen atom in QED at finite $T$ and in~\cite{Brambilla:2011sg} for the case of gluo-dissociation of heavy quarkonium in QCD. Eventually the determination of the thermal rates reduces to evaluating the dipole matrix elements. 

Finally, we comment on the coupling constant. Since in the fundamental theory~\eqref{lag_mod_0} the dark photons couple only to the dark matter fermions, the coupling runs with one flavor at scales larger than $M$, while it is frozen at the value $\alpha = \alpha(M)$ at scales below $M$. 
Thus the coupling constant appearing in the thermal rates discussed so far is in fact a constant.

\section{Rate equations} 
\label{sec:rateequations}
Having presented the relevant low-energy processes and the corresponding thermal rates, we include them in the dynamical rate equations to derive the DM thermal abundance. Here we rely on coupled semi-classical Boltzmann equations, which under certain circumstances, namely for Hubble rates $H(T)$ much smaller than the bound-state decays, may be written as a single effective rate equation, given by~\cite{Ellis_2015}
 \begin{equation}
    (\partial_t + 3H) n = - \frac{1}{2}\langle \sigma_{\textrm{eff}} \, v_{\textrm{rel}} \rangle (n^2-n^2_{\textrm{eq}}) \, .
     \label{Boltzmann_eq_eff}
 \end{equation}
The effective cross section, thermally averaged over the velocities of the incoming unbound pair, is given by\footnote{We perform a thermal average in terms of the Maxwell-Boltzmann distribution of the dark fermions, $f_{\hbox{\tiny MB}}(v_{\hbox{\scriptsize rel}})=\sqrt{2/\pi} \left(M/(2T) \right)^{3/2}
v_{\hbox{\scriptsize rel }}^2  \exp{[-Mv^2_{\textrm{rel}}/(4T)]}$.}
 \begin{equation}
    \langle  \sigma_{\textrm{eff}} \, v_{\textrm{rel}} \rangle  =  \langle \sigma_{\textrm{ann}} v_{\textrm{rel}} \rangle + \sum_{n} \langle   \sigma^n_{\textrm{bsf}} \, v_{\textrm{rel}} \rangle \, \frac{\Gamma_{\textrm{ann}}^n}{\Gamma_{\textrm{ann}}^n+\Gamma_{\textrm{bsd}}^n} \, .
    \label{Cross_section_eff}
 \end{equation}
Here $\Gamma_{\textrm{ann}}^n$ is meant to be replaced by~\eqref{ann_para} or~\eqref{ann_ortho} when performing the summation over S-wave bound states. Equation~\eqref{Cross_section_eff} holds in the limit when bound state-to-bound state transitions are much smaller than $\Gamma_{\textrm{ann}}^n$ and $\Gamma_{\textrm{bsd}}^n$, and may be neglected, which is the so-called no-transition limit. Otherwise it has to be replaced by a more general expression presented in ref.~\cite{Garny:2021qsr}.

\begin{figure}[t!]
    \centering
    \includegraphics[scale=0.86]{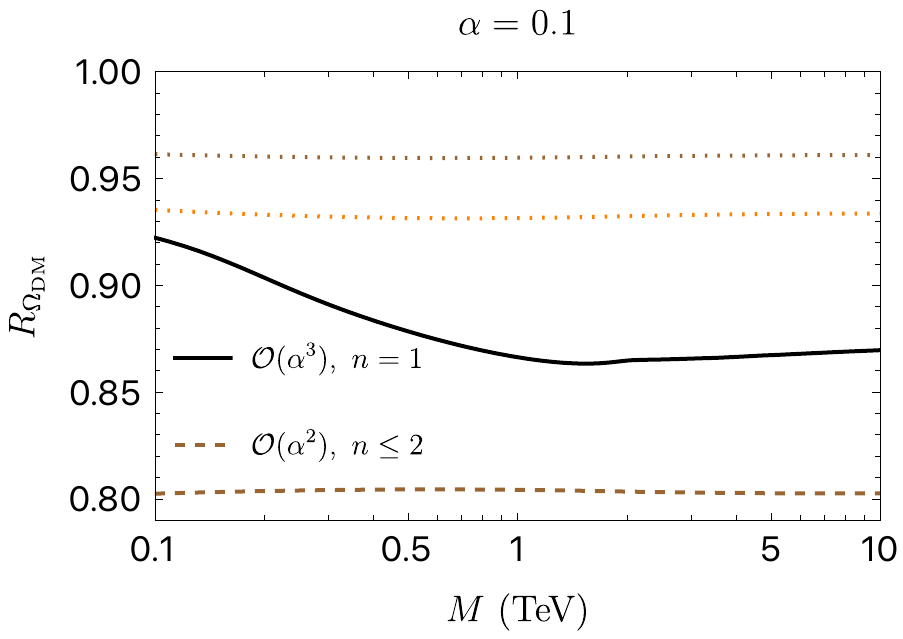}
    \caption{Ratios of different DM energy densities with respect to the energy density obtained from eq.~\eqref{Boltzmann_eq_eff} and~\eqref{Cross_section_eff} for $n=1$ at $\mathcal{O}(\alpha^2)$ in the matching coefficient~\eqref{Im_ds_NLO}, plotted as a function of the DM mass $M$. (Black line) Ratio for the effective cross section for $n=1$ at $\mathcal{O}(\alpha^3)$. (Brown-dotted line) Ratio for the effective cross section for $n=2\textrm{S}$ at order $\alpha^2$ in the matching coefficient~\eqref{Im_ds_NLO}, in the no-transition limit. (Orange-dotted line) Ratio for the effective cross section for $n=10\textrm{S}$ and matching coefficient at $\mathcal{O}(\alpha^2)$, in the no-transition limit. (Brown-dashed line) Ratio for the effective cross section for $n=2$, but neglecting P-wave annihilation widths, at $\mathcal{O}(\alpha^2)$ in~\eqref{pNREFT_2}, beyond the no-transition limit approximation. We recall that the uncertainty in the measured relic density is 1\%. Results are given for $\alpha=0.1$.}
    \label{fig:NLO_versus_nS}
\end{figure}

First, we solve the effective Boltzmann equation~\eqref{Boltzmann_eq_eff} numerically with the effective cross section in~\eqref{Cross_section_eff} for the DM pairs being in the ground state and with the decay width at leading order (LO) in the matching coefficient~\eqref{Im_ds_NLO}, i.e. just given by the paradarkonium width $\Gamma^{1\textrm{S},\hbox{\scriptsize para}}_{\textrm{ann}}=M\alpha^5/2$. This provides our reference energy density. Then, we solve the Boltzmann equation when including the 2S state in the no-transition limit, i.e. considering~\eqref{Cross_section_eff} up to $n=2\textrm{S}$, but still with the paradarkonium decay width at LO. The ratio of the two energy densities is given in fig.~\ref{fig:NLO_versus_nS} by the brown-dotted line. Next, we include S-wave excited states up to 10S. In fig.~\ref{fig:NLO_versus_nS}, we plot the ratio of the obtained energy density with respect to the reference one as the orange-dotted line.

Furthermore, we consider the effective cross section beyond the no-transition limit approximation, i.e., as given in ref.~\cite{Garny:2021qsr}, and solve the Boltzmann equation for the ground state including up to $n=2$ states, but neglecting P-wave annihilation widths, and at order $\alpha^2$ in the matching coefficient~\eqref{Im_ds_NLO}. The ratio with respect to the reference energy density is shown as the brown-dashed curve in fig.~\ref{fig:NLO_versus_nS}, where we see that 2P-to-1S transitions affect the energy density more drastically than just including nS-states in the no-transition limit. 

Finally, we evaluate the impact of $\mathcal{O}(\alpha^3)$ corrections on the 1S state.  We include such corrections in the ground state annihilation width and in $\langle \sigma_{\textrm{ann}} v_{\textrm{rel}} \rangle$. The black solid line in fig.~\ref{fig:NLO_versus_nS} shows the ratio of the obtained energy density with respect to our reference density. The $\mathcal{O}(\alpha^3)$ corrections in the matching coefficients result in a much larger effective cross section due to the additional annihilation channel in the orthodarkonium states. It therefore decreases the DM abundance quite significantly by about 14\% for DM with mass of 1TeV and coupling $\alpha=0.1$.

\section{Conclusions and Outlook}
\label{sec:concl}
In this proceeding, we have summarized the findings of the recent work~\cite{B_and_NAG_1}, where we use the language of NRQED and pNRQED to describe the evolution of thermalized heavy DM pairs in the early universe. Under the hierarchy of scales~\eqref{scale_arrang}, in the EFT dubbed pNRQED$_{\textrm{DM}}$~\eqref{pNREFT_1}, we compute the relevant thermal rates appearing in the evolution equations. We observe that for fermionic DM pairs the additional annihilation in the spin-triplet channel, starting at order $\alpha^3$ in the matching coefficients, gives large contributions to the effective cross section. Furthermore, also bound state-to-bound state transitions appear to play an important role when determining the relic density of DM. Finally, we remark that for large $T$ the multipole expansion may break down, an issue that should be addressed in future works.

\section*{Acknowledgements}
The work of S.B. is supported by the Swiss National Science Foundation under the Ambizione grant PZ00P2\_185783. N.B., G.Q. and A.V. acknowledge support from the DFG cluster of excellence “ORIGINS” under Germany’s Excellence Strategy - EXC-2094-390783311. G.Q. is grateful to the organizers of the NuDM-2022 conference in Egypt for the opportunity to present this research work. The  authors  declare  that  there  are  no  conflicts  of  interest  regarding the publication of this paper. 
%\newpage
\bibliographystyle{JHEP.bst}
\bibliography{LHEP_proceeding.bib}

\providecommand{\href}[2]{#2}\begingroup\raggedright\begin{thebibliography}{10}

\bibitem{Akrami:2018vks}
{\scshape Planck} collaboration, \emph{{Planck 2018 results. I. Overview and
  the cosmological legacy of Planck}},
  \href{https://arxiv.org/abs/1807.06205}{{\ttfamily 1807.06205}}.

\bibitem{Bertone:2004pz}
G.~Bertone, D.~Hooper and J.~Silk, \emph{{Particle dark matter: Evidence,
  candidates and constraints}},
  \href{https://doi.org/10.1016/j.physrep.2004.08.031}{\emph{Phys. Rept.}
  {\bfseries 405} (2005) 279}
  [\href{https://arxiv.org/abs/hep-ph/0404175}{{\ttfamily hep-ph/0404175}}].

\bibitem{Feng:2010gw}
J.~L. Feng, \emph{{Dark Matter Candidates from Particle Physics and Methods of
  Detection}},
  \href{https://doi.org/10.1146/annurev-astro-082708-101659}{\emph{Ann. Rev.
  Astron. Astrophys.} {\bfseries 48} (2010) 495}
  [\href{https://arxiv.org/abs/1003.0904}{{\ttfamily 1003.0904}}].

\bibitem{B_and_NAG_1}
S.~Biondini, N.~Brambilla, G.~Qerimi and A.~Vairo, \emph{{Effective Field
  Theories for Dark Matter Pairs in the Early Universe: cross sections and
  widths}},  \href{https://arxiv.org/abs/2304.00113}{{\ttfamily 2304.00113}}.

\bibitem{Holdom:1985ag}
B.~Holdom, \emph{{Two U(1)'s and Epsilon Charge Shifts}},
  \href{https://doi.org/10.1016/0370-2693(86)91377-8}{\emph{Phys. Lett. B}
  {\bfseries 166} (1986) 196}.

\bibitem{Foot:1991kb}
R.~Foot and X.-G. He, \emph{{Comment on Z Z-prime mixing in extended gauge
  theories}}, \href{https://doi.org/10.1016/0370-2693(91)90901-2}{\emph{Phys.
  Lett. B} {\bfseries 267} (1991) 509}.

\bibitem{Caswell:1985ui}
W.~E. Caswell and G.~P. Lepage, \emph{{Effective Lagrangians for Bound State
  Problems in QED, QCD, and Other Field Theories}},
  \href{https://doi.org/10.1016/0370-2693(86)91297-9}{\emph{Phys. Lett.}
  {\bfseries 167B} (1986) 437}.

\bibitem{Barbieri:1979be}
R.~Barbieri, E.~d'Emilio, G.~Curci and E.~Remiddi, \emph{{Strong Radiative
  Corrections to Annihilations of Quarkonia in QCD}},
  \href{https://doi.org/10.1016/0550-3213(79)90047-6}{\emph{Nucl. Phys. B}
  {\bfseries 154} (1979) 535}.

\bibitem{Hagiwara:1980nv}
K.~Hagiwara, C.~B. Kim and T.~Yoshino, \emph{{Hadronic Decay Rate of Ground
  State Paraquarkonia in Quantum Chromodynamics}},
  \href{https://doi.org/10.1016/0550-3213(81)90181-4}{\emph{Nucl. Phys. B}
  {\bfseries 177} (1981) 461}.

\bibitem{Pineda_1998}
A.~Pineda and J.~Soto, \emph{{Potential NRQED: The positronium case}},
  \href{https://doi.org/10.1103/physrevd.59.016005}{\emph{Phys. Rev. D}
  {\bfseries 59} (1998) }
  [\href{https://arxiv.org/abs/hep-ph/9805424}{{\ttfamily hep-ph/9805424}}].

\bibitem{Pineda:1997ie}
A.~Pineda and J.~Soto, \emph{{The Lamb shift in dimensional regularization}},
  \href{https://doi.org/10.1016/S0370-2693(97)01537-2}{\emph{Phys. Lett.}
  {\bfseries B420} (1998) 391}
  [\href{https://arxiv.org/abs/hep-ph/9711292}{{\ttfamily hep-ph/9711292}}].

\bibitem{Yao:2018nmy}
X.~Yao and T.~Mehen, \emph{{Quarkonium in-medium transport equation derived
  from first principles}},
  \href{https://doi.org/10.1103/PhysRevD.99.096028}{\emph{Phys. Rev. D}
  {\bfseries 99} (2019) 096028}
  [\href{https://arxiv.org/abs/1811.07027}{{\ttfamily 1811.07027}}].

\bibitem{Escobedo:2008sy}
M.~A. Escobedo and J.~Soto, \emph{{Non-relativistic bound states at finite
  temperature (I): The Hydrogen atom}},
  \href{https://doi.org/10.1103/PhysRevA.78.032520}{\emph{Phys. Rev.}
  {\bfseries A78} (2008) 032520}
  [\href{https://arxiv.org/abs/0804.0691}{{\ttfamily 0804.0691}}].

\bibitem{Escobedo:2010tu}
M.~A. Escobedo and J.~Soto, \emph{{Non-relativistic bound states at finite
  temperature (II): the muonic hydrogen}},
  \href{https://doi.org/10.1103/PhysRevA.82.042506}{\emph{Phys. Rev.}
  {\bfseries A82} (2010) 042506}
  [\href{https://arxiv.org/abs/1008.0254}{{\ttfamily 1008.0254}}].

\bibitem{Biondini:2021ccr}
S.~Biondini and V.~Shtabovenko, \emph{{Non-relativistic and potential
  non-relativistic effective field theories for scalar mediators}},
  \href{https://doi.org/10.1007/JHEP08(2021)114}{\emph{JHEP} {\bfseries 08}
  (2021) 114} [\href{https://arxiv.org/abs/2106.06472}{{\ttfamily
  2106.06472}}].

\bibitem{Biondini:2021ycj}
S.~Biondini and V.~Shtabovenko, \emph{{Bound-state formation, dissociation and
  decays of darkonium with potential non-relativistic Yukawa theory for scalar
  and pseudoscalar mediators}},
  \href{https://doi.org/10.1007/JHEP03(2022)172}{\emph{JHEP} {\bfseries 03}
  (2022) 172} [\href{https://arxiv.org/abs/2112.10145}{{\ttfamily
  2112.10145}}].

\bibitem{Brambilla:1999xf}
N.~Brambilla, A.~Pineda, J.~Soto and A.~Vairo, \emph{{Potential NRQCD: An
  Effective theory for heavy quarkonium}},
  \href{https://doi.org/10.1016/S0550-3213(99)00693-8}{\emph{Nucl. Phys.}
  {\bfseries B566} (2000) 275}
  [\href{https://arxiv.org/abs/hep-ph/9907240}{{\ttfamily hep-ph/9907240}}].

\bibitem{Brambilla:2004jw}
N.~Brambilla, A.~Pineda, J.~Soto and A.~Vairo, \emph{{Effective field theories
  for heavy quarkonium}},
  \href{https://doi.org/10.1103/RevModPhys.77.1423}{\emph{Rev. Mod. Phys.}
  {\bfseries 77} (2005) 1423}
  [\href{https://arxiv.org/abs/hep-ph/0410047}{{\ttfamily hep-ph/0410047}}].

\bibitem{Sommerfeld}
A.~Sommerfeld, \emph{{Über die Beugung und Bremsung der Elektronen}},
  {\emph{Ann. Phys.(1931)} {\bfseries 403} (1931) }.

\bibitem{Cassel:2009wt}
S.~Cassel, \emph{{Sommerfeld factor for arbitrary partial wave processes}},
  \href{https://doi.org/10.1088/0954-3899/37/10/105009}{\emph{J. Phys.}
  {\bfseries G37} (2010) 105009}
  [\href{https://arxiv.org/abs/0903.5307}{{\ttfamily 0903.5307}}].

\bibitem{Petraki:2015hla}
K.~Petraki, M.~Postma and M.~Wiechers, \emph{{Dark-matter bound states from
  Feynman diagrams}},
  \href{https://doi.org/10.1007/JHEP06(2015)128}{\emph{JHEP} {\bfseries 06}
  (2015) 128} [\href{https://arxiv.org/abs/1505.00109}{{\ttfamily
  1505.00109}}].

\bibitem{Binder:2020efn}
T.~Binder, B.~Blobel, J.~Harz and K.~Mukaida, \emph{{Dark Matter bound-state
  formation at higher order: a non-equilibrium quantum field theory approach}},
  \href{https://doi.org/10.1007/JHEP09(2020)086}{\emph{JHEP} {\bfseries 09}
  (2020) 086} [\href{https://arxiv.org/abs/2002.07145}{{\ttfamily
  2002.07145}}].

\bibitem{Brambilla:2011sg}
N.~Brambilla, M.~A. Escobedo, J.~Ghiglieri and A.~Vairo, \emph{{Thermal width
  and gluo-dissociation of quarkonium in pNRQCD}},
  \href{https://doi.org/10.1007/JHEP12(2011)116}{\emph{JHEP} {\bfseries 12}
  (2011) 116} [\href{https://arxiv.org/abs/1109.5826}{{\ttfamily 1109.5826}}].

\bibitem{Ellis_2015}
J.~Ellis, F.~Luo and K.~A. Olive, \emph{{Gluino coannihilation revisited}},
  \href{https://doi.org/10.1007/JHEP09(2015)127}{\emph{JHEP} {\bfseries 09}
  (2015) 127} [\href{https://arxiv.org/abs/1503.07142}{{\ttfamily
  1503.07142}}].

\bibitem{Garny:2021qsr}
M.~Garny and J.~Heisig, \emph{{Bound-state effects on dark matter
  coannihilation: Pushing the boundaries of conversion-driven freeze-out}},
  \href{https://doi.org/10.1103/PhysRevD.105.055004}{\emph{Phys. Rev. D}
  {\bfseries 105} (2022) 055004}
  [\href{https://arxiv.org/abs/2112.01499}{{\ttfamily 2112.01499}}].

\end{thebibliography}\endgroup

\end{document}